# A General Framework for the Parametrization of Hierarchical Models

**Omiros Papaspiliopoulos, Gareth O. Roberts and Martin Sköld**

*Abstract.* In this paper, we describe centering and noncentering methodology as complementary techniques for use in parametrization of broad classes of hierarchical models, with a view to the construction of effective MCMC algorithms for exploring posterior distributions from these models. We give a clear qualitative understanding as to when centering and noncentering work well, and introduce theory concerning the convergence time complexity of Gibbs samplers using centered and noncentered parametrizations. We give general recipes for the construction of noncentered parametrizations, including an auxiliary variable technique called the state-space expansion technique. We also describe partially noncentered methods, and demonstrate their use in constructing robust Gibbs sampler algorithms whose convergence properties are not overly sensitive to the data.

## 1. INTRODUCTION

It has long been recognized that the parametrization of a hierarchical model can be crucial for the performance of the Gibbs sampler or the Metropolis–Hastings algorithm which is used to infer about it, and interesting research has been done in that direction. Parametrization methodology is well developed for Gaussian and generalized linear mixed models.

*Omiros Papaspiliopoulos is Assistant Professor, Department of Statistics, University of Warwick, Coventry, CV4 7AL, United Kingdom e-mail: o.papaspiliopoulos@warwick.ac.uk. Gareth O. Roberts is Professor of Statistics, Department of Mathematics and Statistics, Fylde College, Lancaster University, Lancaster LA1 4YF, United Kingdom e-mail: g.o.roberts@lancaster.ac.uk. Martin Sköld is Research Associate, Centre for Mathematical Sciences, Lund University, Box 188, 221 00 Lund, Sweden e-mail: martins@maths.lth.se.*

Research in this area has benefited from the availability of analytic convergence rates for the Gibbs sampler on Gaussian target distributions, provided by [37]. However, the development of general parametrization strategies for nonlinear hierarchical models and in particular for models which involve unobserved (latent) stochastic processes has not received analogous attention; for example, it is listed as one of the important directions for future research in the centennial review by Gelfand [12]. While it is clearly natural and interpretable to specify models according to a hierarchical formulation, more general parametrizations are required for the design of appropriate Markov chain Monte Carlo (MCMC) algorithms for posterior investigation.

Nevertheless, increasingly complex hierarchical models are being used in several areas of applied statistics, including econometrics, geostatistics and genetics, raising high computational challenges. Much research has been devoted to designing high-tech MCMC algorithms that aim to optimize performance within a particular class of models. Such strategies are often ad hoc and not always applicable in a wider framework. Moreover, the bag of MCMC tools and tricks has over the years grown to such an extent that its full power is largely available only to specialists.

Hence, our aim in this paper has been to focus on a *general* strategy that applies to a wide range of statistical contexts. To allow this generality, we may sometimes have to sacrifice the quest for *optimality* for that of *robustness*. Our proposed strategy, when employing MCMC to perform inference for a given hierarchical model, is to use the Gibbs sampler or a similar component-wise updating Metropolis–Hastings algorithm, and to consider two different parametrizations of the model, which we term the *centered* and the *noncentered* parametrizations. The former is usually the default option, and has been used extensively in the literature. The latter, as we argue, turns out to be its natural complement, in the sense that the Gibbs sampler, when under the one parametrization, converges slowly; under the other it often converges much faster. This complementary role of the two parametrizations constitutes the first main advantage of the centered/noncentered framework. The second advantage is the ability to identify, before running any computer program, which of the parametrizations is preferable just by looking at the model structure. In this paper, following a tutorial style we show how certain model structures interact with the efficiency of the Gibbs sampler which is used to infer about them, under both parametrizations. We classify these model structures into three categories, and we demonstrate our arguments on a series of test models, which correspond to popular models used in econometrics, geostatistics and classification. Our test models are presented in their most basic form, since our aim is to present the general methods and issues rather than to solve specific problems. However, detailed references are provided, to direct the interested reader to papers which have dealt in detail with MCMC computation for the more elaborate versions of the models.

## 2. AUGMENTATION SCHEMES AND PARAMETRIZATIONS

In the following we will assume that we are given a set of observed data $Y$, unknown parameters $\Theta$ with associated prior density $P(\Theta)$, and an assumed model $P(Y \mid \Theta)$, which can be conveniently expressed hierarchically using a hidden *layer* $X$ as

$$P(Y \mid \Theta) = \int P(Y \mid X, \Theta) P(X \mid \Theta) \, d\mu(X),$$

where $\mu$ is the measure with respect to which the density $P(X \mid \Theta)$ is defined. Obviously, there are unlimited choices of $X$ which lead to the same posterior inference $P(\Theta \mid Y)$.

In this paper we are concerned with the problem where MCMC methods are employed to get samples from the joint posterior $P(X, \Theta \mid Y)$.

There are several reasons for introducing such an $X$: (i) the *observed likelihood* $P(Y \mid \Theta)$ might be analytically unavailable or too complicated to make inference about $\Theta$, (ii) $X$ itself might have some scientific interpretation and be of statistical interest, and (iii) MCMC methods might be much more efficiently applied to $P(X, \Theta \mid Y)$ than to $P(\Theta \mid Y)$.

We define a (*re*)*parametrization* of an augmentation scheme $X$, by any random pair $(X^*, \Theta)$ with joint prior density $P(X^*, \Theta)$ (with respect to an arbitrary dominating measure) together with a function $h$ such that

(2.1) $$X = h(X^*, \Theta, Y).$$

In this general setup, $h$ need not be one-to-one. We call a reparametrization *practical* if $P(\Theta \mid X^*)$, hence $P(\Theta \mid X^*, Y)$, is known up to a normalizing constant.

Once a parametrization $(X^*, \Theta)$ is chosen, MCMC is used to obtain samples from the joint posterior $P(X^*, \Theta \mid Y)$. These samples can be easily transformed [using (2.1)] to yield samples from $P(X, \Theta \mid Y)$. The purpose of the reparametrization in this framework is solely to improve the efficiency of the Monte Carlo method.

Hence, for a particular model, we have the dual choice of an augmentation scheme and a parametrization. This paper is concerned with efficient choice of the latter, since this issue is more amenable to a *general* discussion than the former.

EXAMPLE 2.1 (*Random effects model*). As a simple example, we consider the simple linear Gaussian random effects model,

$$Y_i \sim \mathrm{N}(C_i X_i, \Sigma_y),$$
$$X_i \sim \mathrm{N}(D\Theta, \Sigma_x), \quad \text{independently for } i = 1, \dots, n,$$

where $C_i$ are individual-level covariates, $X_i$ are individual-level regression parameters, centered around population parameters $\Theta$, and $\Sigma_y, \Sigma_x$ are covariance matrices. In this case $X = \{X_1, \dots, X_n\}$, $Y = \{Y_1, \dots, Y_n\}$, and a family of practical reparametrizations is described by the transformations

$$X_i = X_i^* + A_i,$$

$1 \le i \le n$, for vectors $A = \{A_i, 1 \le i \le n\}$ which are allowed to depend only on $\theta$. Remarkably, even when $A$ is restricted to depend only linearly on $\theta$, this family of reparameterizations is sufficiently flexible to include complete posterior orthogonalization of the state space of $(X, \Theta)$.

In this paper, we shall not focus on reparametrization methods for $\Theta$, which may of course be important for the construction of efficient MCMC methods.

### 2.1 Hierarchically Centered Parametrizations

In this article, we will base the discussion around a particular form of augmentation scheme described by the graphical model in Figure 1. The corresponding parametrization of the model in terms of $(X, \Theta)$ is commonly known as the centered parametrization (CP). This terminology was first introduced in the context of generalized linear hierarchical models by Gelfand, Sahu and Carlin [13, 14], and later in a more general context by Papaspiliopoulos, Roberts and Sköld [33] as the parametrization under which the data $Y$ are independent of the parameters $\Theta$ conditionally on the imputed data $X$. The hierarchical model structure in Figure 1 is common in statistical practice and crops up in very diverse areas.

Example 2.1 is an example of a hierarchically centered parametrization and is one of the simplest possible models with graphical structure given in Figure 1. Here we give some further "canonical" examples, in their simplest form, in order to illustrate the diversity of the areas for which the methods we present in this paper are relevant.

EXAMPLE 2.2 (*Geostatistical model*). In this case $X = \{X(u), u \in B \subset \mathbf{R}^2\}$ is an unobserved spatial field, assumed to be a realization from a stochastic process, for example, a Gaussian process with mean and covariance functions specified up to a few unknown parameters $\Theta$ (see, e.g., [9]). For example, one could take $\mu$ to be the mean level (the specification could easily be adapted to include covariates), and $\sigma^2 \Sigma(\alpha)$ to be the covariance function, where $\sigma^2$ is the overall variance and $\Sigma(\alpha)$ is a positive-definite correlation matrix with parameter $\alpha$ which controls the range of spatial dependence. In this case $\Theta = (\mu, \sigma, \alpha)$. The data consist of conditionally independent observations $Y_i \sim f(Y_i \mid X(u_i))$, where $f$ is some density, at a finite number of locations, $u_i$, $i = 1, \ldots, n$; thus $Y = \{Y_1, \ldots, Y_n\}$.

EXAMPLE 2.3 (*Diffusion stochastic volatility model*). Diffusion processes are used extensively in financial mathematics and econometrics for modeling the time-evolution of stock prices and exchange rates. An empirically observed characteristic of financial time-series data is that the instantaneous variance is time varying and exhibits a stochastic behavior. A popular class of diffusion-based models which can describe this behavior is the stochastic volatility models [5, 15], which have the following hierarchical structure:

$$dY_t = \exp(X_t/2) \, dB_{1,t}, \qquad t \in [0, T],$$
$$dX_t = b(X_t) \, dt + \Theta \, dB_{2,t},$$

where $B_{j,t}$, $j = 1, 2$, are standard Brownian motions. In this case $Y_t$ is a continuous-time process, for example, the log-price of the exchange rate between two currencies. $X = \{X_t, t \in [0, T]\}$ represents the time-varying log-variance of $Y_t$, which is also assumed to be a diffusion process, parametrized in terms of $b(\cdot)$ (which we assume to be completely known to simplify things) and $\Theta$, which controls the variation in the $X$ process. The process $Y_t$ is observed at time instances $t_i$, $i = 1, \ldots, n$, $Y = \{Y_{t_1}, \ldots, Y_{t_n}\}$.

### 2.2 Convergence of the Gibbs Sampler Under Different Parametrizations

When used to sample from $P(X^*, \Theta \mid Y)$, the algorithm starts with arbitrary $\Theta^0$, and iterates

(2.2)
$j.1$ Update $X^{*j}$ from the distribution of $X^* \mid \Theta^{j-1}, Y$.

$j.2$ Update $\Theta^j$ from the distribution of $\Theta \mid X^{*j}, Y$.

for $j = 1, \ldots, N$. The output $\Theta^1, \ldots, \Theta^N$ is now a dependent and approximate sample from $P(\Theta \mid Y)$. The problem that the sample is approximate is an important one; see, for example, [19] for a recent review and synthesis of the methods available for eliminating the so-called burn-in problem. However, the major weakness of MCMC is that the serial dependence can be very strong within the Monte Carlo sample, rendering it unreliable for posterior inference, since it will give estimates with very high (or even infinite) variance.

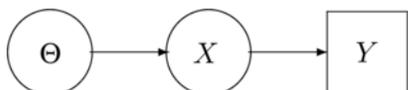

FIG. 1. *Graphical model of the centered parametrization (CP).*

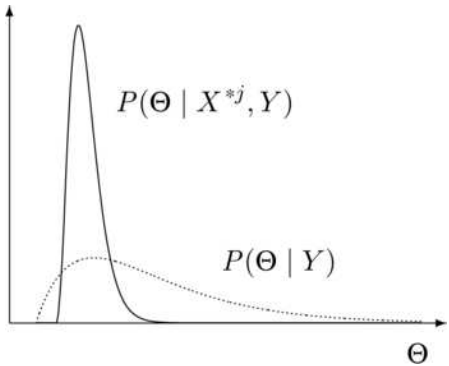

FIG. 2. *Posterior distributions of $\Theta$ given observed (dotted) and full augmented (solid) data for a particular value $X^{*j}$ of imputed data $X^*$.*

A useful heuristic interpretation of the (lack of) performance of Gibbs sampling schemes like (2.2) is given in Figure 2. An efficient algorithm here is one where the support of $P(\Theta \mid X^{*j}, Y)$ closely mimics that of the target $P(\Theta \mid Y)$ and performance is closely related to the relative sizes of the typical distance traveled in a single update [i.e., the "size" of the support of $P(\Theta \mid X^{*j}, Y)$] and the distance we need to travel to cover the support of $P(\Theta \mid Y)$ (i.e., the "size" of this support). This heuristic can be neatly formalized using the concept of the *Bayesian fraction of missing information* (e.g., [23, 39]), defined for a real-valued, square-integrable function of interest $f(\Theta)$ as

$$(2.3) \quad \gamma_f = 1 - \frac{E(\text{Var}(f(\Theta) \mid X^*, Y) \mid Y)}{\text{Var}(f(\Theta) \mid Y)}.$$

Liu [23] shows that the *maximal correlation coefficient*, defined as

$$(2.4) \quad \gamma = \sup_f \gamma_f,$$

where the supremum is taken over all square-integrable functions $f$, is the geometric convergence rate of the Gibbs sampler under the $(X^*, \Theta)$ parametrization. He also shows (see also [1, 24]) that

$$(2.5) \quad \begin{aligned} \gamma &= \sup_f \text{Corr}(f(\Theta^j), f(\Theta^{j+1})) \\ &= \left\{ \sup_{f,g} \text{Corr}(f(\Theta), g(X^*) \mid Y) \right\}^2, \end{aligned}$$

where $\Theta^i, \Theta^{i+1}$ are consecutive output values of the Gibbs sampler started at stationary, $\Theta^0 \sim P(\Theta \mid Y)$.

Thus, values of $\gamma$ close to 1 correspond to slowly mixing algorithms, resulting in high autocorrelation in the sampled $\Theta$ values, and are caused by high dependence between the updated components. Here "close to 1" should be interpreted as *very* close: an algorithm with $\gamma = 0.9$ or even $0.99$ mixes sufficiently well for most practical purposes. Hence, as suggested in the Introduction, we will not aim for $\gamma \approx 0$ but rather try to provide the user with an alternative for those (surprisingly common) "*very* close to 1" situations. It is also important to realize that $\gamma$ corresponds to the worst-case scenario; it is the rate at which the slowest-mixing functionals converge to their stationary means. For geometrically ergodic algorithms, a more intuitive measure of efficiency is what we will call the *mixing time*, and denote by $\tau$,

$$\tau = -\frac{1}{\log \gamma} \approx \frac{1}{1 - \gamma} \quad \text{for } \gamma \approx 1.$$

$\tau$ is proportional to the number of iterations needed for the Markov chain to be within certain (total variation, say) distance from stationarity.

As we have mentioned, the centered is often the default parametrization and we will refer to the Gibbs sampler under CP [i.e., (2.2) with $X^* = X$] as the centered algorithm (CA), and denote by $\gamma_c$ and $\tau_c$ its convergence rate and mixing time, respectively. There are a number of reasons why a CA forms a natural starting point. Conveniently, the conditional independence structure in Figure 1 implies that the conditional posterior $P(\Theta \mid X, Y) = P(\Theta \mid X)$ is often easy to sample from. Moreover, the analysis in [13, 14] showed that the CA is very efficient for location parameters in (generalized) linear mixed models.

Nevertheless, in view of (2.5) the potential drawbacks of hierarchical centering are easily understood. Note that $X$ and $\Theta$ are generally strongly dependent a priori, and to diminish this dependence the data $Y$ need to be strongly informative about $X$; for a simple illustration see Figure 3. Section 3, however, reveals important situations where the data, even when they are informative about $\Theta$, cannot diminish the prior dependence between $X$ and $\Theta$. In particular, in Section 3 we demonstrate that CA: (1) can have a mixing time which increases as the sample size increases, for example, when the latent model $P(X \mid \Theta)$ fails to satisfy classical regularity conditions (Section 3.1); (2) can have very unstable behavior (e.g., be nongeometrically ergodic), due to certain robustness properties of the model [in this case the maximal correlation coefficient is not appropriate to quantify the dependence between $X$ and $\Theta$ (Section 3.2)]; and (3) can be reducible due to the choice of the augmentation scheme, for example, when the observed data are finite dimensional but the imputed data are infinite dimensional, as in Example 2.3 (Section 3.3).

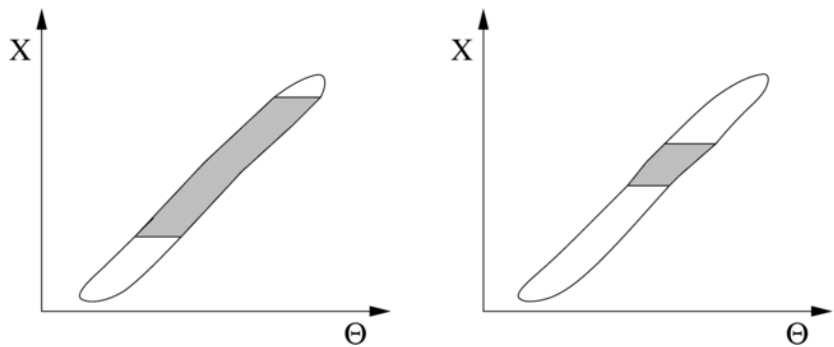

FIG. 3. *Schematic figure depicting the region of highest support of prior (unshaded) and posterior (shaded) distributions, for the CP where Y is weakly (left) and strongly (right) informative about X. This illustrates how strongly informative data break down prior dependence between $\Theta$ and X, whereas weakly informative data do not.*

One remedy to these problems is to use the Gibbs sampler under a reparametrization. An important—and as we show in this paper, complementary to CP—subclass of reparametrizations $X^*$ are the *hierarchically noncentered parametrizations* (NCPs). The name originates from normal linear hierarchical models [13], but Papaspiliopoulos [30] gave the general definition, that a reparametrization $X^*$ is called *noncentered* when $X^*$ and $\Theta$ are *a priori independent*. It is often straightforward for a given model to find a noncentered parametrization, as Section 3 shows, but sometimes we need to resort to more involved techniques, as we discuss in Section 4.

As an example, we return to Example 2.1, and note that the parametrization which takes

$$\tilde{X}_i = X_i - D\Theta$$

trivially satisfies the requirements of being an NCP. In more complex problems, identifying NCPs is typically much harder.

Returning to the general case, the graphical model corresponding to this parametrization is given in Figure 4. Of course, the conditions of a *practical* reparametrization are trivially satisfied since by construction $P(\Theta \mid X^*) = P(\Theta)$ is available up to a normalizing constant and hence we can use complex transformations $h$ without having to worry about computing Jacobian terms. To distinguish this class of parametrizations we will denote them by $\tilde{X}$ rather than $X^*$. We will also use the notation NCA when referring to the Gibbs sampler under NCP, and $\gamma_{nc}, \tau_{nc}$ to refer to its convergence rate and mixing time, respectively.

Since $\tilde{X}$ and $\Theta$ are a priori independent, the NCA can be much more efficient than the CA when $X$ is relatively (to $\Theta$) weakly identified by the data. In lieu, when $X$ is well identified, $\tilde{X}$ and $\Theta$ are often a posteriori strongly dependent by means of (2.1); an illustration is given in Figure 5.

In this paper, we consider CP and NCP as *general* classes of *practical* parametrizations. Neither is "optimal" in any reasonable sense, and clearly for any particular problem there might be parametrizations which outperform (in terms of the efficiency of the corresponding Gibbs sampler) both of them. However, the merit of the centered/noncentered algorithms lies in that we can say very general things about their performance regardless of the specificities of the particular model they are applied to. Their strength lies in their ease of extension to high-dimensional parameter spaces and complicated data structures. Moreover CA and NCA are complementary, in the sense that *NCAs are likely to perform well when the CAs do not* and conversely, as we show in Section 3. For all these reasons, as we will show in the paper, in several cases we can predict their performance just by examining the model at hand.

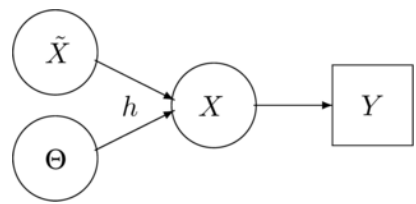

FIG. 4. *Graphical model of NCP.*

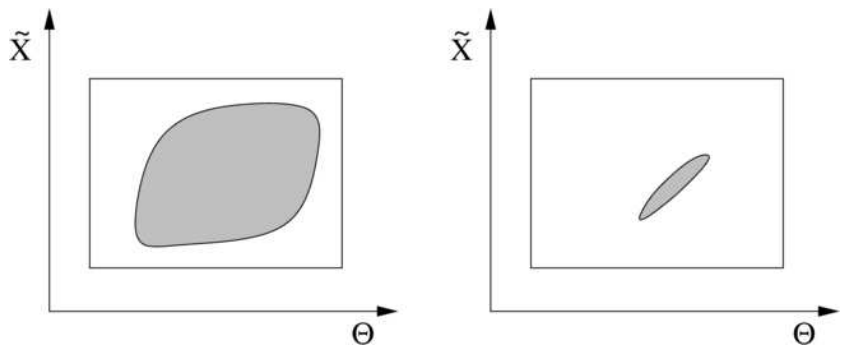

FIG. 5. *Schematic figure of prior (unshaded) and posterior (shaded) probability regions for the NCP where $Y$ is weakly (left) and strongly (right) informative about $X$ (not $\tilde{X}$). Here strong data induce strong posterior dependence between $\Theta$ and $\tilde{X}$. Contrast the panels with the corresponding in Figure 3.*

It is worth mentioning at this stage that the convergence rate analysis we have presented in this section does not directly apply to algorithms which update either of the components $X^*$ or $\Theta$ using a Metropolis–Hastings step. This updating is often done in complex models, where drawing directly from the typically multivariate distribution $P(X^*|\Theta, Y)$ is not feasible. However, the intuition that the higher the posterior dependence between $X^*$ and $\Theta$ the slower the convergence can be useful even for these more general algorithms.

## 3. CASE STUDIES

We here give a few stylized examples to illustrate the importance of the choice of centering/noncentering and their complementary roles. The examples are divided into three subcategories, corresponding to models that are likely to occur in practice and where choice of parametrization can be crucial. Bear in mind that the examples are stripped down to their most basic versions for illustrative purposes. To avoid losing the big picture we use simple heuristic arguments; however, all the claims can be proved relatively easily. The notation adopted below clarifies how NCP is constructed.

### 3.1 Models where the Efficiency Depends on Sample Size

An important characteristic of any MCMC algorithm is how its efficiency scales with the size of the data set.

EXAMPLE 3.1 (*Repeated measurements*). Our first example is in the flavor of Gelfand, Sahu and Carlin [13], and Example 2.1. Consider the simple hierarchical model described by

$$Y_i \sim N(X, \sigma_y^2), \quad \text{independently for } i = 1, \ldots, n,$$
$$X = \tilde{X} + \Theta, \quad \tilde{X} \sim N(0, \sigma_x^2),$$

where $P(\Theta) \propto 1$ is chosen. Here $(\Theta, X)$ is the CP and $(\Theta, \tilde{X})$ is the NCP.

The fact that the model is Gaussian implies that the supremum in (2.4) is attained by linear functionals; thus $\tau_c = O(1/\log n)$. This is similar to the asymptotic argument Gelfand, Sahu and Carlin [13] used to support the use of the CA for linear models. On the other hand, $\tau_{nc} = O(n)$. Hence, the CA improves while the NCA deteriorates with sample size for this simple model.

EXAMPLE 3.2 (*Hidden Markov model*). Consider the following trivial hidden Markov model (HMM):

$$Y_i \sim N(X_i, \sigma_y^2), \quad \text{independently for } i = 1, \ldots, n,$$
$$X_i = \tilde{X}_i + \Theta,$$

where $\tilde{X}_i$ follows Markov dynamics with stationary distribution $N(0, \sigma_x^2)$, for instance through $\tilde{X}_i = \rho \tilde{X}_{i-1} + (1-\rho^2)^{1/2} \sigma_x Z_i$ for some appropriate standard normal innovation $Z_i$ and constant $|\rho| < 1$. Also assume that the proper prior $P(\Theta) \equiv N(0, 1)$ is chosen. This time both $\tau_c$ and $\tau_{nc}$ are $O(1)$.

In the case where $\tilde{X}_i$ are chosen to be independent, a complete analytic solution is available for the Gibbs sampler (from [37]). For instance, the presence of the data to "tie down" the hidden states is crucial for the performance of CA. In the absence of data, that is, when we want to sample from the prior, $\tau_{nc} = O(1)$ (trivially since NCA delivers i.i.d.

samples from the prior), but now $\tau_c = O(n)$ and performance of CA will deteriorate with $n$. We note that the assumption of conditional independence in the hidden states is not crucial (see Example 3.3 though). Both algorithms will be $O(1)$ if the $X$ process is modeled to have moderate serial dependence (the NCP needs to be defined appropriately in this case); see, for example, [33].

The assumption of Gaussian errors in the observation and hidden equation is a very important one. The remaining examples in this section, and those in Section 3.2, elaborate on these non-Gaussian extensions.

EXAMPLE 3.3 (*Nonregular parameters*). The following example illustrates how CA can deteriorate with $n$ even if the number of random effects and data are equal ($n$). Nonregular in this example refers to the problem of estimating $\Theta$ given $X$. Assume the hierarchical model

$$Y_i \sim N(X_i, 1), \quad \text{independently for } i = 1, \ldots, n,$$
$$X_i = \Theta \tilde{X}_i, \qquad \tilde{X}_i \sim \text{Uniform}(0, 1),$$

where $P(\Theta) \propto 1$ is chosen. What makes this model particular is that $\text{Var}(\Theta \mid X) = O(n^{-2})$ while $\text{Var}(\Theta \mid Y) = O(n^{-1})$ and hence $\tau_c \geq O(n)$. On the other hand, it can be shown that $\tau_{nc} = O(1)$.

A similar problem with CA can occur in a variety of situations where the latent model $P(X \mid \Theta)$ fails to satisfy classical regularity conditions, including, for example, latent autoregressions $X_i = \Theta X_{i-1} + \varepsilon_i$ for Gaussian innovations in the explosive case $\Theta > 1$ [45] and Exponential innovations in general [28].

In the above example CA ran into problems due to the changing support of $P(X \mid \Theta)$ with $\Theta$. Below, we give two examples where instead the support of $P(X \mid \Theta, Y)$ changes with $Y$ and the relative performances of CA and NCA are reversed.

EXAMPLE 3.4 (*Stochastic frontier models*). This is a class of models very popular in economics (see, e.g., [17]), which in their simplest form are written as

$$Y_i = X_i - u_i,$$
$$u_i \sim \text{Exp}(\lambda), \quad \text{independently for } i = 1, \ldots, n,$$
$$X_i = \Theta + \tilde{X}_i, \quad \tilde{X}_i \sim N(0, \sigma_x^2),$$

where the $u_i$'s represent company-specific inefficiencies, and we again choose $P(\Theta) \propto 1$. What makes this model challenging from a computational perspective is the presence of the asymmetric error in the observation equation. This does not cause problems to CA, since it is easy to see that $\tau_c = O(1)$. On the contrary, since $\Theta \mid \tilde{X}, Y \sim \text{Exp}(n\lambda)$ truncated on $\Theta > \max_i \{Y_i - \tilde{X}_i\}$, $\text{Var}(\Theta \mid \tilde{X}, Y) = (n\lambda)^{-2}$; thus $\tau_{nc} \geq O(n)$. Similar models are also used for reaction-time data; see, for example, [34].

EXAMPLE 3.5 (*Rounded data*). Here we consider models for truncated data. Assume for example that

$$Y_i = \lfloor X_i \rfloor, \qquad \text{independently for } i = 1, \ldots, n,$$
$$X_i = \tilde{X}_i + \Theta, \quad \tilde{X}_i \sim N(0, \sigma_x^2),$$

where we use the notation $\lfloor x \rfloor$ to denote the largest integer less than or equal to $x$.

Under CP the augmented and the observed information are of the same $O(n)$; thus $\tau_c = O(1)$. On the other hand, $P(\Theta \mid \tilde{X}, Y)$ is supported in $[\max_i \{Y_i - \tilde{X}_i\}, \min_i \{Y_i + 1 - \tilde{X}_i\}]$; hence $\text{Var}(\Theta \mid \tilde{X}, Y) = O(n^{-2})$ and $\tau_{nc} = O(n)$.

EXAMPLE 3.6 (*Bayesian classification*). Consider the problem of Bayesian classification as in, for example, Lavine and West [22]. Assume we are interested in whether a set of covariates $Y$ can be used to distinguish between two classes of individuals (e.g., healthy/sick). Based on a training sample with known classes, we have derived the posterior predictive distributions $f_0$ and $f_1$ of the covariates based on the respective classes. The same covariates are now measured on a number $n$ of individuals with unknown status. Hence, we have independent data $Y_1, \ldots, Y_n$ from the two-component mixture model with density $\Theta f_0(Y_i) + (1 - \Theta) f_1(Y_i)$, where $\Theta$ is the proportion of individuals of class 0 to which we assign a uniform prior. For computational convenience, this model is usually augmented with indicators $X_i$ and rewritten in hierarchical form as (see, e.g., [8])

$$Y_i \sim f_{X_i}(Y_i), \qquad \text{independently for } i = 1, \ldots, n,$$
$$X_i = \mathbf{1}\{\tilde{X}_i \leq \Theta\}, \quad \tilde{X}_i \sim \text{Uniform}(0, 1).$$

Here we might consider two extreme scenarios (with $N = \sum_{i=1}^n X_i$):

1. Perfect classification. If $f_0$ and $f_1$ have disjoint supports, we can perfectly distinguish the classes and $P(\Theta \mid X, Y) = P(\Theta \mid Y)$. Hence, the CA will produce independent replications from the posterior. For the NCA, however, $\Theta \mid \tilde{X}, Y \sim \text{Uniform}(\tilde{X}_{(N)}, \tilde{X}_{(N+1)})$. Hence $\text{Var}(\Theta \mid \tilde{X}, Y) = O(n^{-2})$ and $\tau_{nc} > O(n)$ since $\text{Var}(\Theta \mid Y) = O(n^{-1})$.

2. *Completely unrelated covariates.* Now assume the covariates turn out not to be related to class, that is, $f_0 = f_1$. Here $P(\Theta \mid \tilde{X}, Y) = P(\Theta)$ and the NCA will produce independent replications. On the other hand, $\Theta \mid X \sim \text{Beta}(N+1, (n-N)+1)$ and $\text{Var}(\Theta \mid X) = O(n^{-1})$ while $\text{Var}(\Theta \mid Y) = O(1)$. Hence, $\tau_c > O(n)$.

Of course realistic examples often tend to fall in between scenarios 1 and 2 above. Nevertheless, they convey some intuition about the behavior in their neighborhoods.

### 3.2 Models where the Efficiency Depends on the Tails of the Links

Section 2.1 related the convergence rate of the two-component Gibbs sampler to the maximal correlation coefficient $\gamma$ between the updated components. Nevertheless, $\gamma$ is a measure of *global dependence*; it is thus inappropriate when the dependence between $X$ and $\Theta$ is very different in different parts of the space. In these situations, CA can have qualitatively very different behavior than NCA. The following characteristic example is taken from [32].

EXAMPLE 3.7 (*Heavy-tailed HMM*). Consider the following "innocent"-looking modification of the HMM given in Example 3.2:

$Y_i \sim \text{Cauchy}(X_i, \sigma_y)$,

independently for $i = 1, \ldots, n$,

$X_i = \tilde{X}_i + \Theta, \quad \tilde{X}_i \sim \text{N}(0, \sigma_x^2)$,

where the improper prior $P(\Theta) \propto 1$ is chosen. Therefore, we have *robustified* the observation equation by replacing the Gaussian error by a Cauchy with median $X_i$ and scale parameter $\sigma_y$. Figure 6 shows the contours of the posterior distribution of $(X, \Theta)$, for $n = 1$, $Y = 0$, $\sigma_y = 1$, $\sigma_x^2 = 5$. Near the mode, $X$ and $\Theta$ are approximately independent: however, as $\Theta \to \infty$, the conditional posterior distribution of $X \mid \Theta, Y$ tends to be concentrated around $\Theta$ and ignores the data $Y$. The reason for this is that when $|\Theta - Y|$ is large, the two sources of information about $X$, the prior and the likelihood, are conflicting with each other, and inference for $X$ is dominated by the link with the lighter tails. In fact, Papaspiliopoulos and Roberts [32] show that CA is not even geometrically ergodic (therefore $\gamma_c = 1$), whereas NCA is geometrically (in fact it is uniformly) ergodic. Note that in the complementary model

$Y_i \sim \text{N}(X_i, \sigma_y^2)$, independently for $i = 1, \ldots, n$,

$X_i = \tilde{X}_i + \Theta, \quad \tilde{X}_i \sim \text{Cauchy}(0, \sigma_x)$,

the performance of CA and NCA is reversed, and CA is uniformly ergodic while NCA is subgeometric.

Papaspiliopoulos and Roberts [32] derive very general results for hierarchical models with linear structure, including time-series and spatial models where $X$ is a latent Gaussian field which is observed at certain locations with heavy-tailed noise. Models with heavy-tailed observation error are used for protecting the inference about $X$ from outlying observations; see, for example, [3, 20, 43].

### 3.3 Models where the Efficiency Depends on the Amount of Imputation

In modern complex hierarchical modeling, it is common to impute infinite-dimensional objects $X$, or alternatively very fine finite-dimensional approximations of them. The canonical example in this case is the diffusion stochastic volatility model, presented in Example 2.3 and revisited in Example 3.8 below, but this type of augmentation scheme is undertaken also in Bayesian nonparametric modeling, as we show in Example 3.9 below. Often in such situations *ergodicity constraints* link $X$ and $\Theta$, in the sense that $\Theta$ (or some components of it, if it is a vector) might be completely specified given $X$. Then, the convergence of CA deteriorates as the approximation gets finer (it is reducible in the limit). One approach to circumvent this problem is to jointly update $\Theta$ and $X$, advocated, for example, in Darren Wilkinson's discussion to [33] and in [21]. However, joint updates are often complicated to implement, and might, for example, necessitate reversible-jump algorithms that are difficult to tune and diagnose. Here we argue that NCA often provides a simpler and efficient alternative in such situations. The examples are natural opposites to Example 3.1 in the sense that dimension of *augmented* rather than *observed* data increases with $n$.

EXAMPLE 3.8 (*Diffusion stochastic volatility model, revisited*). Here we revisit the model presented in Example 2.3. Assume for illustration that one observation from the price process is available, $Y = Y_{t_1}$, thus $Y \mid X \sim \text{N}(0, \int_0^{t_1} e^{X_s} \, ds)$. However, the distribution of the path integral is not known; thus in order to compute the likelihood one needs to impute the whole path $X = \{X_s, 0 \le s \le t_1\}$, or as it is usually done, a very fine discretization $X_{1:n} = \{X_{it_1/n}, i = 0, \ldots, n\}$, for large $n$, and approximate the integral appropriately. However, the *quadratic*

*variation identity*

$$\lim_{n\to\infty} \sum_{i=1}^{n} (X_{it_1/n} - X_{(i-1)t_1/n})^2 = \Theta^2$$

shows that $\Theta$ is completely determined by any string of sample path $X$ ($n = \infty$), whereas for large but finite $n$, $X_{1:n}$ and $\Theta$ are very strongly dependent a priori. The limiting centered algorithm which imputes $X$ is actually reducible: for any starting $\Theta^0$, $P(X \mid Y, \Theta^0)$ is a distribution on the path space concentrated on paths for which their quadratic variation is $(\Theta^0)^2$; and conditionally on such $X^1 \sim P(X \mid Y, \Theta^0)$, $P(\Theta \mid Y, X^1)$ is concentrated on $\Theta^0$. Roberts and Stramer [38] show that the CA on $(X_{1:n}, \Theta)$ is $O(n)$; that is, the better we try to approximate the likelihood [and therefore the posterior $P(\Theta \mid Y)$], the worse the convergence of the algorithm.

A possible choice for NCP is to take $\tilde{X} = \{B_{2,s}, 0 \leq s \leq t_1\}$, the driving Brownian motion of the volatility equation, and the associated approximation $\tilde{X}_{1:n} = \{B_{2,it_1/n}, i = 0, \ldots, n\}$; this approach has recently been adopted in [5]. NCA is $O(1)$, thus it does not deteriorate as the approximation to the true continuous-time model gets better. Roberts and Stramer [38] suggested an alternative reparametrization $X^*$, which qualifies to be called noncentered only when $b = 0$ (a case, however, which has motivated the construction). If we define $X_t^* = X_t/\Theta$, then direct application of Îto's formula shows that

$$dX_t^* = \frac{b(\Theta X_t^*)}{\Theta} dt + dB_{2,t}.$$

Clearly, when $b = 0$, $X^*$ is an NCP and in fact coincides with the one suggested earlier. When $b \neq 0$, the distribution of $X^*$ does depend on $\Theta$; however, any realization of $X^*$ contains only finite information about $\Theta$. (Diffusion sample paths on a fixed time interval, say $[0, t_1]$, under mild regularity conditions contain finite information about any parameters in the drift, but infinite information about parameters in the diffusion coefficient.) Therefore, the Gibbs algorithm which updates $\Theta$ and a fine approximation $X^*_{1:n} = \{X^*_{it_1/n}, i = 0, \ldots, n\}$ will also be $O(1)$. In the context of diffusion processes, this parametrization is particularly appealing, as it will be demonstrated in Section 5.

EXAMPLE 3.9 (*Bayesian nonparametrics*). A very active area of research is that of Bayesian nonpara-metric statistics. In this area, unknown distributions and functions are modeled using flexible stochastic process priors. For example, a popular choice is the Dirichlet process prior [11], which is a prior on the space of all distribution functions on a given space, say $\mathcal{Z}$. When we write that $P \sim DP(\Theta, H)$, we imply that $P$ is a random measure distributed according to the Dirichlet process prior with parameters $\Theta > 0$, and $H$ which is a (known) distribution on $\mathcal{Z}$. Many methods for carrying out inference on models containing hidden Dirichlet components work by integrating out the Dirichlet process. In what follows we will deliberately not adopt this approach, with a view toward generalizations to other nonparametric models where integrating out is not an option.

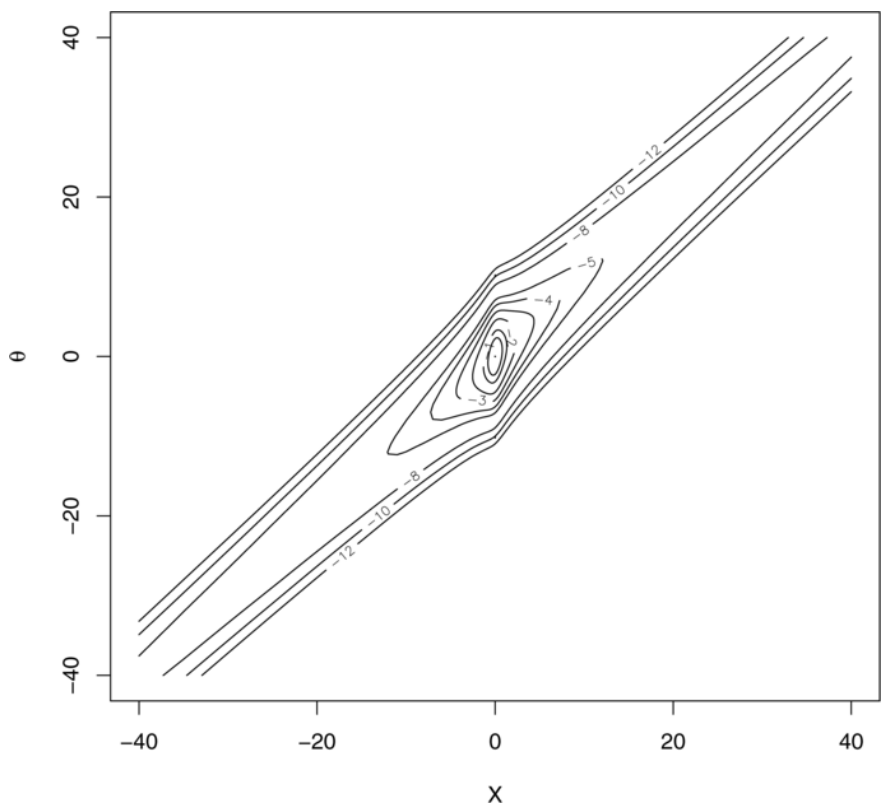

FIG. 6. *Contour plot of the posterior distribution of $(X, \Theta)$ from the heavy-tailed HMM, with $n = 1$, $Y = 0$, $\sigma_y = 1$, $\sigma_x^2 = 5$. Notice that near the mode, $X$ and $\Theta$ are roughly independent, whereas they become highly dependent in the tails.*

A constructive series representation [42] gives that

$$(3.1) \qquad P(\cdot) = \sum_{i=1}^{\infty} p_i \delta_{Z_i}(\cdot),$$

where $Z_i$ are i.i.d. variables from $H$, and

$p_1 = X_1$ and

$p_i = (1 - X_1)(1 - X_2) \cdots (1 - X_{i-1}) X_i$, for $i \geq 2$,

where the $X_i$'s are independent Beta$(1, \Theta)$ variables. Therefore, every realization of $P$ is an infinite mixture of random point masses $p_i$, located at random locations $Z_i \in \mathcal{Z}$. Notice that $X$ and $\Theta$ are linked through the *ergodicity constraint*

$$\lim_{n \to \infty} -\frac{1}{n} \sum_{i=1}^{n} \log(1 - X_i) = \Theta.$$

$P$ is commonly used at an intermediate stage in hierarchical models. A popular class of such models, used for classification, density estimation and nonparametric regression, is the Dirichlet mixture models; see, for example, [10, 26]. When $X = \{X_1, X_2, \ldots\}$ and $Z = \{Z_1, Z_2, \ldots\}$ are imputed in order to facilitate inference for the hierarchical model, as for example in [31], the CA is reducible since $X$ and $\Theta$ are linked by the constraint given above. Approximate centered methods, which impute only an approximation $X_{1:n} = \{X_1, \ldots, X_n\}, Z_{1:n} = \{Z_1, \ldots, Z_n\}$, as in [18], will have $O(n)$ mixing time. Papaspiliopoulos and Roberts [31] suggest a simple NCP to come round this problem, given by the inverse-CDF construction, $X_i = 1 - (\tilde{X}_i)^{1/\Theta}$ for i.i.d. uniforms $\tilde{X}_i \sim$ Uniform$(0, 1)$.

Similar computational problems appear in a variety of other nonparametric models, where, for example, Lévy processes are used at intermediate stages in hierarchical models. For a description of such models, see, for example, [44].

## 4. CONSTRUCTING NCP'S

We have already promoted the centered/noncentered methodology as a default strategy for tackling challenging computational problems, and we have given insight on when each should be used (or avoided) through the case studies in Section 3. Typically, CP is naturally suggested by the augmentation scheme chosen for a particular hierarchical model (in fact usually they coincide). On the other hand, in the examples we have presented so far NCP has been quite obvious (maybe with the exception of the diffusion model in Example 3.8). Some tricks we have applied were:

- Location: if $X \sim F(\cdot - \Theta)$,
  then $h(\tilde{X}, \Theta) = \tilde{X} + \Theta$, $\tilde{X} \sim F(\cdot)$ (e.g., Examples 3.2, 3.7).
- Scale: if $X \sim F(\cdot/\Theta)$,
  then $h(\tilde{X}, \Theta) = \Theta \tilde{X}$, $\tilde{X} \sim F(\cdot)$ (e.g., Examples 3.3, 3.8).
- Inverse CDF: if $X \sim F_\Theta(\cdot)$ where $F_\Theta$ is a distribution function on $\mathbf{R}$,
  then $h(\tilde{X}, \Theta) = F_\Theta^{-1}(\tilde{X})$, $\tilde{X} \sim$ Uniform$(0, 1)$ (Examples 3.6, 3.9).

Clearly, the above "recipes" can be combined in different ways. For example, suppose that $X$ is a latent Gaussian field as in Example 2.2. An NCP that has been successfully applied in this context (see, e.g., [25] in the context of log-Gaussian Cox processes and [7] for spatial generalized linear mixed models) is

$$(4.1) \qquad X = h(\tilde{X}, \Theta) = \sigma \Sigma(\alpha)^{1/2} \tilde{X} + \mu,$$

for a vector of i.i.d. standard normals $\tilde{X}$ and where $\Sigma(\alpha)^{1/2}$ is, for example, provided by a Cholesky factorization of $\Sigma(\alpha)$. Similarly, if $X_1, \ldots, X_t$ is a Markov chain with transition CDF $F_\Theta(\cdot \mid \cdot)$, the inverse CDF can be applied recursively to construct an NCP through $X_i = F_\Theta^{-1}(\tilde{X}_i \mid X_{i-1})$ for a sequence of independent uniforms $\tilde{X}_1, \ldots, \tilde{X}_n$.

It is not easy to give very general instructions on how to produce an NCP for arbitrary latent structures. In a way, this is the price to pay for having a method which *in principle* could be applied in any context. However, several probabilistic relationships can be exploited; see, for example, [30]. Generally, if you know how to simulate from the prior $P(X \mid \Theta)$, a careful inspection of the algorithm often naturally suggests an NCP. Indeed, the tricks displayed above are all standard tools in simulation. Noncentering of stochastic processes can often be achieved by reference to various representation theorems. For example, we followed this approach in Example 3.9 using a constructive representation of the Dirichlet process.

An important observation, however, which can be used to construct NCP for complex latent structures is that $\tilde{X}$ can take values on a much higher-dimensional space than $X$ and the function $h$ in (2.1) can be noninvertible, as we discuss in the sequel.

### 4.1 State-Space Expansion

Notice that the Gibbs sampler which samples from the posterior $P(\tilde{X}, \Theta \mid Y)$ can be implemented in the following manner. Assume that the current state is $(X^{j-1}, \Theta^{j-1})$:

$j$.1 Update $X$ according to the distribution of
$X \mid \Theta^{j-1}, Y$, to give $X^{j-1/2}$.

$j$.2 Transform $(X^{j-1/2}, \Theta^{j-1}) \to \tilde{X}^{j-1/2}$.

$j$.3 Update $\Theta^j$ according to the distribution of
$\Theta \mid \tilde{X}^{j-1/2}, Y$.

$j$.4 Transform $(\tilde{X}^{j-1/2}, \Theta^j) \to X^j$.

Here step $j$.4 is unnecessary if it is possible to sample directly from the distribution in $j$.1. We have included steps $j$.1–2 and $j$.4, rather than a single step updating $\tilde{X}^j$ from $\tilde{X} \mid \Theta^{j-1}, Y$, for two reasons: first, it illustrates how code from a scheme like (2.2) could be reused, and second, it allows for situations where $\tilde{X} \mid \Theta^{j-1}, Y$ is not easily updated.

While step $j$.2 in (4.2) above will often be implemented through a deterministic transformation $\tilde{X} = h^{-1}(X, \Theta)$, equality in distribution is sufficient. Hence, the step can be performed by a draw from $\tilde{X} \mid X^j, \Theta^{j-1}$, which allows for situations where $h$ is a many-to-one function. In Example 3.6, for instance, where $X = \mathbf{1}\{\tilde{X} \leq \Theta\}$, in step $j$.2 we draw $\tilde{X}$ from a Uniform$(\Theta, 1)$ if $X = 0$ and from a Uniform$(0, \Theta)$ distribution if $X = 1$.

Another idea, suggested and elaborated in [30], is that step $j$.3 in (4.2) can sometimes be performed using only partial information about $\tilde{X}$, and hence the full $\tilde{X}$ need not be computed in step $j$.2. This is the case when $j$.3 is performed through an accept/reject mechanism like the Metropolis–Hastings algorithm and $h(\tilde{X}, \cdot)$ only needs to be computed for the current value $\Theta^{j-1}$ of $\Theta$ and a proposed new value $\Theta'$. Specifically, $\tilde{X}$ can live on an infinite-dimensional space (while the dimension of $X$ can still be finite) and this flexibility makes noncentering feasible to apply in very general contexts; see the example below for an illustration and [36] for a fully worked example. The method described in Example 4.1 below has recently been found necessary in order to construct MCMC algorithms for inference for discretely observed diffusion processes in [2].

EXAMPLE 4.1 (*Latent Poisson processes*). Poisson processes are natural a priori models for event times and spatial locations of objects. Often data are partial or consist of a noisy image from which spatial locations ($X$) and intensity ($\Theta$) are to be inferred. Note that in latent point process models, $X \mid \Theta, Y$ can rarely be updated exactly and slow reversible-jump algorithms are commonly applied. Hence, it is especially important here that the MCMC algorithm can move freely along the $\Theta$-axis.

We will discuss two possible ways to construct an NCP $\tilde{X}$ for a Poisson process $X$ with rate $\Theta$ on the unit interval. They both involve state-space expanding techniques (see, e.g., [30, 33]) where $\tilde{X}$ is formally an infinite-dimensional object. The ideas carry over in a straightforward manner to Poisson processes on arbitrary spaces with general intensity functions parametrized in terms of a vector of parameters $\Theta$:

1. $\tilde{X}$ is a unit rate Po-process on $[0,1] \times [0, \infty)$. If $\{(\tilde{u}_1, \tilde{v}_1), \ldots\}$ is an enumeration of the point-coordinates of $\tilde{X}$, we define $X = h(\tilde{X}, \Theta) = \{\tilde{u}_i; \tilde{v}_i \leq \Theta\}$ (see the left frame in Figure 7).
2. $\tilde{X}$ is a unit rate Po-process on $[0, \infty)$ with points $\{\tilde{u}_1, \ldots\}$ and $X = h(\tilde{X}, \Theta) = \{\tilde{u}_i/\Theta; \tilde{u}_i \leq \Theta\}$.

Again, what makes the above strategies feasible is that for a fixed $\Theta$, $X$ and hence $L(h(\tilde{X}, \Theta), \Theta)$ only depends on a finite subset of the points in $\tilde{X}$. Hence, if $\Theta$ is updated using a Metropolis–Hastings algorithm, we only need to store partial information about $\tilde{X}$ in each step. More implementation details with an application to shot-noise type stochastic volatility processes can be found in [36].

EXAMPLE 4.2 (*Hidden Markov jump processes*). Given the above recipes for constructing NCPs for Poisson processes, the step to Markov jump processes is not far. Such models are natural for describing the variation of populations and epidemics in time; see, for example, [4, 27, 29, 35] for a variety of applications of hidden Markov jump processes.

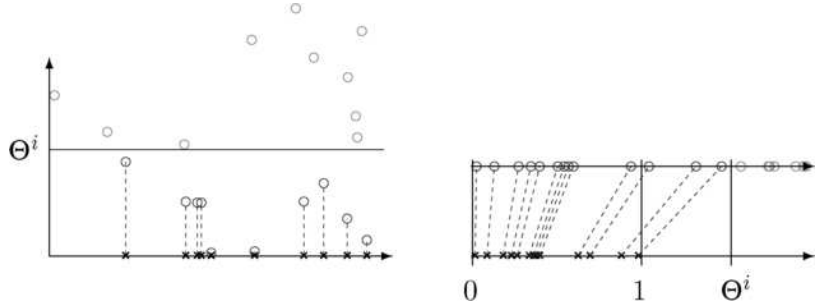

Fig. 7. *Two different NCPs of a $\Theta$-rate Poisson process on the unit interval. In the left figure $\tilde{X}$ is a unit rate Po-process on $[0,1] \times [0,\infty)$ and in the right on $[0,\infty)$. Circles denote the points of $\tilde{X}$ and crosses the points of $X$ obtained after the many-to-one transformation $X = h(\tilde{X}, \Theta)$.*

For example, we might consider an immigration-death process $X(t)$, $t \in [0,T]$, defined through

$$P(X(t+h) = k \mid X(t) = l)$$
$$= \lambda h + o(h), \quad k = l+1,$$
$$= 1 - (\lambda + l\mu)h + o(h), \quad l = l,$$
$$= l\mu h + o(h), \quad k = l-1,$$
$$= o(h), \quad \text{otherwise,}$$

for $i = 0, 1, \ldots$ and with $\Theta = (\lambda, \mu)$ for immigration and death rates $\lambda$ and $\mu$, respectively.

Processes like $X(t)$ are often simulated recursively based on a sequence of independent exponential and uniform variables, corresponding to interarrival times and type of event, respectively. Hence, we define $\tilde{X}_i = (\tilde{z}_i, \tilde{u}_i)$, $i = 1, \ldots$, with $\tilde{z}_i$ and $\tilde{u}_i$ being $\text{Exp}(1)$ and $\text{Uniform}(0,1)$, respectively. Further, denoting by $\tau_i$ the time of the $i$th event, the transformation $X = h(\tilde{X}, \Theta)$ is defined recursively through

$$\tau_i = \tau_{i-1} + \tilde{z}_i/(\lambda + X(\tau_{i-1})\mu),$$
$$X(t) = X(\tau_{i-1}) + 2\mathbf{1}\{\tilde{u}_i \leq \lambda/(\lambda + X(\tau_{i-1})\mu)\} - 1,$$
$$t \in [\tau_{i-1}, \tau_i),$$

for $\tau_0 = 0$, $i = 1, \ldots$.

## 5. ROBUSTNESS AND DATA-BASED MODIFICATIONS

When implementing "automatic" MCMC routines in software one is often faced with the problem that, under the same model, performance can vary between different data sets (see, e.g., [6]). Hence, for example, CP might be preferred by some data sets and NCP by others, depending on how informative the particular realization of the data is for $X$. Another problem is that the relative performance of the two strategies might vary within the support of $P(X, \Theta \mid Y)$. We have already encountered this problem in Example 3.7 (see especially Figure 6); here a CP is preferable around the mode while it is necessary to use an NCP in the tails.

Fortunately, it is not necessary to make a choice between a CA and an NCA: alternating the two within the same algorithm might greatly improve the robustness without causing increase in computational time of practical relevance. In fact, such a combined strategy has been shown to be necessary in a number of practical situations (e.g., [16, 27, 32]).

### 5.1 Correcting for the Presence of Data

Both CPs and NCPs are constructed exclusively based on the prior distributions of the model. Since this prior distribution is usually manageable (in comparison with the posterior), we have seen how both parametrizations can be applied through a wide class of statistical models and data structures. Inevitably, the a priori construction can also be viewed as a weakness and one might argue that when efficiently parametrizing the posterior, data should be taken into account.

In this section we will discuss models where simple data-based modifications of CPs and NCPs either are necessary or simply improve performance. We show that even when CPs and NCPs are unsuccessful, they often form the first step toward an efficient strategy.

5.1.1 *Correcting the CP.* Problems relating to the CP are fairly easily understood and it is natural to look for reparametrizations in terms of conditionally pivotal quantities $X^*$.

Consider, for example, a linear reparametrization

(5.1) $\qquad X = v^{1/2}(\Theta, Y)X^* + m(\Theta, Y),$

where the linearity in $X^*$ ensures that $P(\Theta \mid X^*)$ is available. If $X$ belongs to a location-scale family, an NCP is given by $v(\Theta, Y) = \text{Var}(X \mid \Theta)$ and $m(\Theta, Y) = \text{E}(X \mid \Theta)$. For more general families this can be seen as a first-order approximation of an NCP. When correcting this parametrization for the presence of data, it is natural to consider $v(\Theta, Y) = \text{Var}(X \mid \Theta, Y)$ and $m(\Theta, Y) = \text{E}(X \mid \Theta, Y)$. Indeed, when also $X \mid \Theta, Y$ is location-scale, this choice will make $X^*$ independent of $\Theta$. The approach allows data to decide the parametrization and naturally gives an NCP for "infinitely weak data" and a CP for "infinitely strong data." Hence, (5.1) can be interpreted as a *partially noncentered* parametrization (see [30, 33]).

While these choices of $v$ and $m$ are not analytically available in general, they can often be sufficiently well approximated from, for example, a quadratic expansion of the log-posterior; see [6] for a successful application in the context of spatial GLMMs.

5.1.2 *Correcting the NCP.* Of course, we can also consider reparametrizations of NCPs, that is,

(5.2) $\quad X = h(\tilde{X}, \Theta) = h(\tilde{h}(X^*, \Theta, Y), \Theta).$

While in (5.1) it was natural to search for an approximate pivotal quantity $X^*$, the function $\tilde{h}$ in (5.2) often fills the purpose of relieving hard constraints on $X$ imposed by data. The situation arises naturally in models for discretely observed stochastic processes, where the unobserved process values are augmented to facilitate likelihood calculations or prediction.

EXAMPLE 5.1 (*Discretely observed diffusions*). This is close in spirit to Examples 2.3 and 3.8. We again assume that $X_t$ is the solution to the stochastic differential equation

$$dX_t = b(X_t)\,dt + \Theta\,d\tilde{X}_t;$$

however, instead of $X$ being unobserved, here it is *discretely* observed. Thus, assuming for simplicity one observation, the data are $Y = X_{t_1}$, for some $0 < t_1$. The likelihood function, $P(X_{t_1} \mid \Theta)$, is typically unavailable when $b(x)$ is a nonlinear function of $x$. Instead, it is easy (using the so-called Girsanov formula) to write down the likelihood given the complete path $X = \{X_s, s \le t_1\}$, or at least to approximate it accurately enough using a fine discretization $X_{1:n} = \{X_{it_1/n}, i = 0, \ldots, n\}$, for large $n$. It can be recognized that the setting is very similar to that in Example 3.8, and we have already seen that a CA which imputes $X_{1:n}$ has $O(n)$ mixing time, due to the quadratic variation identity binding $\Theta$ with $X_{1:n}$, and should be avoided. Nevertheless, the NCA which was successful in Example 3.8 does not work here, since by construction, $Y$ is a deterministic function of $\Theta$ and $\tilde{X}$. Therefore it is impossible to update $\Theta$ given $\tilde{X}$ without violating the constraint imposed by the data.

The natural sequence of reparametrizations for this problem is given by

$$X(t) = \Theta\tilde{X}(t) = \tilde{h}(X^*, \Theta, Y) = \Theta(X^*(t) - tY/\Theta);$$

see [38] for more details.

EXAMPLE 5.2 (*Gaussian Markov random fields*). In many applications, observed data can be represented as a Gaussian field sampled at irregularly spaced locations. A well-known problem with this approach is that likelihood computations can be burdensome, especially for large data sets, due to the necessary inversion of the covariance matrix. On the other hand, if the covariance matrix is sparse, computations can be speeded up considerably. One approach (due to Rue [40]) to allow for fast estimation is to approximate the Gaussian field by a Markov random field, by embedding the points into a rectangular lattice equipped with a (suitable) Markov random field neighborhood structure. As a result, the covariance matrix $\Sigma$ of the approximate model is sparse. The unavailable observations at the empty lattice points are treated as missing data and are imputed. For notational simplicity, we assume that only the location parameter $\Theta$ is unknown. Moreover, we set $X = (Y_1, \ldots, Y_n, X'_1, \ldots, X'_m)^T$ to be the values of the field at the lattice point, arranged such that the first $n$ values are actually observed, and the remaining $m$ are unobserved. Therefore, under the approximate model, $X \mid \Theta \sim \text{N}(\Theta, \Sigma)$, $(X, \Theta)$ is the CP, and $X$ is a partially observed process. Given the complete data $X$, and assuming a flat prior for $\Theta$, we have that

$$P(\Theta \mid X) \propto \exp(-(X - \Theta\mathbf{1})^T \Sigma^{-1} (X - \Theta\mathbf{1})/2),$$

where $\mathbf{1} = (1, \ldots, 1)^T$.

Often a large number of lattice points needs to be augmented ($n \ll m$) (see, e.g., [41]), hence the augmented data can contain vastly more information about $\Theta$ than observed data $Y = (Y_1, \ldots, Y_n)$ and the CA will perform poorly. On the other hand, the corresponding NCA defined through $\tilde{X} = X - \Theta\mathbf{1} \sim \text{N}(0, \Sigma)$ will in fact be reducible since $\Theta = Y_i - \tilde{X}_i$, $i = 1, \ldots, n$.

Here it is natural to choose $X^* = (X_1^*, \ldots, X_m^*)$ to be zero mean with the appropriate covariance structure and

$$X = \tilde{X} + \Theta \mathbf{1} = \tilde{h}(X^*, \Theta, Y)$$
$$= (Y_1 - \Theta, \ldots, Y_n - \Theta, X_1^*, \ldots, X_m^*)^T + \Theta \mathbf{1}.$$

Hence, in effect we center the observed and noncenter the unobserved values.

## 6. DISCUSSION

We have tried to demonstrate in this paper the very wide applicability and complementary roles of centering and noncentering approaches to parametrization of hierarchical models. Neither centering nor noncentering methods are uniformly effective; however, we give a clear qualitative (and in some cases quantitative) understanding of when noncentering parametrizations are likely to be effective. Conveniently, the two strategies possess complementary strengths, making their combined use an attractive and relatively robust option.

A significant advantage of centered over noncentered methods is that they explicitly use the conditional independence structure inherent from the hierarchical setup. It is common for conjugate conditional distributions for parameters to exist only when using the CP. Thus, in some cases this can lead to a significant relative computational cost for the NCP. Our experience suggests that in many cases, any extra computational burden is more than offset by the convergence advantages resulting from the appropriate use of noncentering. See [27] for a simulation study of algorithm efficiency in the partially observed stochastic epidemic context. This work explicitly considers the loss of efficiency due to the breakdown of conjugacy and compares this with the benefits of more rapid convergence.

Noncentered parametrizations are certainly not unique. We describe how quite general classes of models allow the easy construction of noncentered parametrizations. Perhaps the most generally applicable technique for this purpose is the state-space expansion technique. Here a key feature of noncentered parametrizations is that they are by construction always *practical*, that is, $P(\Theta|\tilde{X}) = P(\Theta)$ is always available up to a normalizing constant.

It is by now well established empirically and in some cases theoretically, that parametrizations which are superior to both the centered and noncentered strategies can usually be constructed, either by constructing a suitable continuum of parametrizations between centered and noncentered, or perhaps by more data-specific methods such as the construction used for diffusion processes in Example 5.1. Partially noncentered parametrizations are sometimes difficult to construct, though the methods have appealing robustness characteristics with respect to different data sets and data types.

Further work is required in a number of directions. Applications of (partial) noncentering parametrizations are as yet uncommon, and much remains to be learned about the effectiveness of the method, particularly taking into account the possible additional computing overheads sometimes associated with the use of these methods. Also further study is required to generalize our theoretical knowledge beyond the examples we are able to cover here.

## ACKNOWLEDGMENTS

The first author would like to thank Mike Pitt for useful suggestions, and EPSRC Grant GR/S61577/01 for financial support. The Editor handling this paper made helpful comments which have improved the presentation of the paper.